\documentclass[showpacs,amssymb,amsmath,aps,prd,secnumarabic,twocolumn,groupedaddress]{revtex4-1}

\usepackage{graphicx}

\begin{document}

\title{Observation of \boldmath{$\chi_c$} and \boldmath{$\chi_b$} nuclear suppression via dilepton polarization measurements}

\author{Pietro Faccioli and Jo\~ao Seixas}

\affiliation{Laborat\'orio de Instrumenta\c{c}\~ao e F\'{\i}sica Experimental
de Part\'{\i}culas (LIP), 1000-149 Lisbon, Portugal, \\
Centro de F\'{\i}sica Te\'orica de Part\'{\i}culas (CFTP), 1049-001 Lisbon, \\
Physics Department, Instituto Superior T\'ecnico (IST), 1049-001 Lisbon,
Portugal}

\date{\today}

\begin{abstract}

We demonstrate that it is possible to use the polarization of vector quarkonia,
measured from dilepton event samples, as an instrument to study the suppression
of $\chi_c$ and $\chi_b$ in heavy-ion collisions, where a direct determination
of signal yields involving the identification of low-energy photons is
essentially impossible. A change of the observed J/$\psi$ and $\Upsilon(1S)$
polarizations from proton-proton to central nucleus-nucleus collisions would
directly reflect differences in the nuclear dissociation patterns of $S$ and
$P$ states and may provide a strong indication for quarkonium sequential
suppression in the quark-gluon plasma.

\end{abstract}



\maketitle


\section{Introduction}
\label{sec:intro}

Hypotheses on the suppression of $\chi_c$ and $\chi_b$ production in
nucleus-nucleus collisions play a crucial role in the interpretation of the
J/$\psi$ and $\Upsilon(1S)$ measurements from SPS~\cite{bib:SPS},
RHIC~\cite{bib:RHIC} and LHC~\cite{bib:LHC} in terms of evidence of quark-gluon
plasma (QGP) formation. The observation of the $\chi_c$ and $\chi_b$
suppression patterns in Pb-Pb collisions at the LHC could confirm or falsify
the ``sequential quarkonium melting'' scenario~\cite{bib:sqm,bib:Karsch} and,
therefore, discriminate between the QGP interpretation and other options.
However, a direct observation of the $\chi_c$ and $\chi_b$ signals in their
radiative decays to ${\rm J}/\psi$ and $\Upsilon(1S)$ is practically impossible
in heavy-ion collisions, given the very large number of background photons
produced in such events.

In this paper we show that this important missing piece of information can be
obtained by measuring how the dilepton decay distributions of prompt J/$\psi$
and $\Upsilon(1S)$ change from proton-proton (or peripheral nucleus-nucleus) to
central nucleus-nucleus collisions, bypassing the difficulty of explicitly
identifying the events containing $\chi_c$ and $\chi_b$ decays.

The demonstration of the result is made by steps. In Sec.~\ref{sec:basics} we
describe how the polarization is transferred in the decays to lighter
quarkonium states and summarize the present knowledge of the feed-down
contributions in collider experiments. In Sec.~\ref{sec:scenarios} we consider
two illustrative polarization scenarios, complementing the existing
experimental information on prompt-J/$\psi$ and $\Upsilon(1S)$ polarizations
with educated guesses about the still unknown polarizations of the $P$-states.
In Sec.~\ref{sec:formulas} we formalize the relations between the
$P$-to-$S$-state feed-down fractions and the observable polarizations of
directly and indirectly produced states. Using these ingredients, in
Sec.~\ref{sec:suppression} we illustrate the main result of the paper,
discussing the feasibility of the $\chi$ suppression measurement using dilepton
angular distributions at the LHC.

The notations used in this paper for angles and polarization parameters are
those defined in Ref.~\cite{bib:EPJC}. We report here for convenience the most
general form of the observable angular distribution of J/$\psi$ and $\Upsilon$
decays into lepton pairs:
\begin{eqnarray}
  W(\cos \vartheta, \varphi)
  & \propto & \, \frac{1}{(3 + \lambda_{\vartheta})} \,
  (1 + \lambda_{\vartheta} \cos^2 \vartheta \label{eq:observable_ang_distr} \\
  & + &  \lambda_{\varphi} \sin^2 \vartheta \cos 2 \varphi +
  \lambda_{\vartheta \varphi} \sin 2 \vartheta \cos \varphi ) \, ,\nonumber
\end{eqnarray}
where $\vartheta$ and $\varphi$ are the polar and azimuthal emission angles of
one of the leptons with respect to a system of axes defined in the dilepton
rest frame and $\lambda_{\vartheta}, \lambda_{\varphi}, \lambda_{\vartheta
\varphi}$ are the anisotropy parameters.

\section{Polarization contributions from feed-down decays}
\label{sec:basics}

Many of the prompt J/$\psi$ and $\Upsilon$ mesons produced in hadronic
collisions result from the decay of heavier $S$- or $P$-wave quarkonia.
However, the existing polarization measurements at collider energies make no
distinction between directly and indirectly produced states. The role of the
feed-down from heavier $S$ states (responsible, for example, for about $8\%$ of
J/$\psi$ production at low $p_{\rm T}$~\cite{bib:feeddown}) is rather well
understood. Data of the BES~\cite{bib:BES_psiprime} and
CLEO~\cite{bib:CLEO_upsilon} experiments in $e^+e^-$ collisions indicate that
in the decays $\psi^{\prime} \rightarrow \mathrm{J}/\psi \pi \pi$ and
$\Upsilon(2S) \rightarrow \Upsilon(1S) \pi \pi$ the di-pion system is produced
predominantly in the spatially isotropic ($S$-wave) configuration, meaning that
no angular momentum is transferred to it. Consequently, the angular momentum
alignment is preserved in the transition from the $2S$ to the $1S$ state. This
allows us to assume that the dilepton decay angular distribution of the
J/$\psi$ [$\Upsilon(1S)$] mesons resulting from $\psi^{\prime}$
[$\Upsilon(2/3S)$] decays is the same as the one of the $\psi^{\prime}$
[$\Upsilon(2/3S)$], provided that a common polarization axis is chosen for the
two particles. At high momentum, when the J/$\psi$ and $\psi^{\prime}$
directions with respect to the centre of mass of the colliding hadrons
practically coincide, $\psi^{\prime}$ mesons and J/$\psi$ mesons from
$\psi^{\prime}$ decays have the same observable polarization with respect to
any system of axes defined on the basis of the directions of the colliding
hadrons. In the case of the polar anisotropy parameter $\lambda_\vartheta$, for
instance, the relative error, $|\Delta\lambda_\vartheta/\lambda_\vartheta|$,
induced by the approximation of considering the J/$\psi$ and $\psi^{\prime}$
directions as coinciding is $\mathcal{O}[(\Delta m / p)^2]$, where $\Delta m$
is the $2S-1S$ mass difference and $p$ the total laboratory momentum of the
dilepton. For $p>5$~GeV/$c$ this error is of order 1\%. Moreover, the directly
produced J/$\psi$ [$\Upsilon(1S)$] and $\psi^\prime$ [$\Upsilon(2/3S)$] are
expected to have the same production mechanisms and, therefore, very similar
polarizations. As a consequence, the polarization of J/$\psi$ [$\Upsilon(1S)$]
from $\psi^\prime$ [$\Upsilon(2/3S)$] can be considered to be almost equal to
the polarization of directly produced J/$\psi$ [$\Upsilon(1S)$], so that, at
least in first approximation, the two contributions can be treated as one.

\begin{table}[t]
\begin{center}
\begin{tabular}{cccc}
\hline
  $J_z$  & $\lambda_\vartheta(S)$ & $\lambda_\vartheta(P_1)$ & $\lambda_\vartheta(P_2)$  \\
\hline
 $0$     &         $-1$           &      $+1$                &     $-3/5$     \\
 $\pm 1$ &         $+1$           &      $-1/3$              &     $-1/3$     \\
 $\pm 2$ &           -            &        -                 &     $+1$       \\
\hline
\end{tabular}
\caption{ Values of the observable polar anisotropy parameter
$\lambda_\vartheta$ of the J/$\psi$ [$\Upsilon(1S)$] dilepton decay
distribution, corresponding to pure angular momentum states ($J_z$) of the
\emph{directly} produced particle (J/$\psi$ [$\Upsilon(1S)$] itself, $\chi_1$
or $\chi_2$). The results are obtained in the high-momentum approximation (for
$p>5$~GeV/$c$ they are affected by an error smaller than $1\%$) and are valid
not only in the E1 approximation (usually assumed in the literature), but also
including all orders of the photon radiation
expansion~\cite{bib:chi_polarization}. } \label{tab:lambda_vs_helicity}
\end{center}
\end{table}

On the contrary, the J/$\psi$ [$\Upsilon(1S)$] mesons resulting from
$\chi_{cJ}$ [$\chi_{bJ}$] radiative decays can have very different
polarizations with respect to the directly produced ones. Directly produced $P$
and $S$ states can originate from different partonic and long-distance
processes, given their different angular momentum and parity properties.
Moreover, the emission of the spin-1 and always transversely polarized photon
necessarily changes the angular momentum projection of the $q \bar{q}$ system
in the $P \to S$ radiative transition. As a result, the relation between the
``spin-alignment'' of the directly produced $P$ or $S$ state and the shape of
the observed dilepton angular distribution is totally different in the two
cases: for example, if directly produced $\chi_{c1}$ and J/$\psi$ both had
``longitudinal'' polarization (angular momentum projection $J_z = 0$ along a
given quantization axis), the shape of the dilepton distribution would be of
the kind $\propto 1 - \cos^2\!\vartheta$ for the direct J/$\psi$ and $\propto 1
+ \cos^2\!\vartheta$ for the J/$\psi$ from $\chi_{c1}$.
Table~\ref{tab:lambda_vs_helicity} lists values of the observable polar
anisotropy parameter $\lambda_\vartheta$ of the J/$\psi$ [$\Upsilon(1S)$]
dilepton decay distribution, corresponding to pure angular momentum states of
the \emph{directly} produced particle. In particular, while for directly
produced $S$ states $-1 < \lambda_\vartheta < +1$, for those from decays of
$P_1$ and $P_2$ states the lower bound is $-1/3$ and $-3/5$, respectively. More
detailed constraints on the three anisotropy parameters $\lambda_\vartheta$,
$\lambda_\varphi$ and $\lambda_{\vartheta \varphi}$ in the cases of directly
produced $S$ state and $S$ states from decays of $P_1$ and $P_2$ states can be
found in Ref.~\cite{bib:chi_polarization}. Figure 3 of that work shows that the
allowed parameter space of the decay anisotropy parameters for the directly
produced J/$\psi$ [$\Upsilon(1S)$] strictly includes the one of the $S$-states
from $P_2$ decays, which, in turn, strictly includes the one of the $S$-states
from $P_1$ decays.

The feed-down fractions are not well-known experimentally. In the charmonium
case, the $\chi_{c}$-to-J/$\psi$ and $\chi_{c2}$-to-$\chi_{c1}$ yield ratios
have been measured by CDF~\cite{bib:cdf_chic} in the rapidity interval $|y|<
0.6$, with insufficient precision to indicate or exclude important $p_{\rm T}$
dependencies. The $p_{\rm T}$-averaged results,
\begin{align}
\begin{split}
R(\chi_{c1}) + R(\chi_{c2}) & = 0.30 \pm 0.06 \, , \\
\quad R(\chi_{c2}) / R(\chi_{c1}) & = 0.40 \pm 0.02 \, , \label{eq:Rchic}
\end{split}
\end{align}
where $R(\chi_{c1})$ and $R(\chi_{c2})$ are the fractions of prompt J/$\psi$
yield due to the radiative decays of $\chi_{c1}$ and $\chi_{c2}$, effectively
correspond to a phase-space region (low $p_{\rm T}$ and central rapidity), much
smaller than the one covered by the LHC experiments.

CDF also measured~\cite{bib:cdf_chib} the fractions of $\Upsilon(1S)$ mesons
coming from radiative decays of $1P$ and $2P$ states as, respectively,
$R(\chi_{b1})+R(\chi_{b2}) = (27 \pm 8)\%$ and $R(\chi_{b1}^\prime) +
R(\chi_{b2}^\prime) = (11 \pm 5)\%$, for $p_{\rm T} > 8$~GeV$/c$ and without
discrimination between the $J=1$ and $J=2$ states. These results tend to
indicate that the contribution of the feed-down from $P$ states to
$\Upsilon(1S)$ production is at least as large as in the corresponding
charmonium case, even if the experimental error is quite large. The same
indication is provided with higher significance by the $\Upsilon$ polarization
measurement of E866~\cite{bib:e866_upsilon}, at low $p_{\rm T}$, as discussed
in the next section.

\section{Two example scenarios}
\label{sec:scenarios}

In this section we derive, using available experimental and theoretical
information, two illustrative scenarios for the polarizations of the charmonium
and bottomonium families. In our considerations we will use the following
addition rule~\cite{bib:invariants}:
\begin{equation}
\vec{\lambda}^\mathrm{prt} \, = \, \frac{ \frac{ [1 - R(P_1) - R(P_2)] \,
\vec{\lambda}^\mathrm{dir} } { 3 + \lambda_\vartheta^\mathrm{dir} } + \frac{
R(P_1) \, \vec{\lambda}^{P_1} } { 3 + \lambda_\vartheta^{P_1} } + \frac{ R(P_2)
\, \vec{\lambda}^{P_2} } { 3 + \lambda_\vartheta^{P_2} }
 } { \frac{ [1 - R(P_1) - R(P_2)] } { 3 + \lambda_\vartheta^\mathrm{dir} } +
\frac{ R(P_1) } { 3 + \lambda_\vartheta^{P_1} } + \frac{ R(P_2) } { 3 +
\lambda_\vartheta^{P_2} }
 } \, ,
\label{eq:addition_rule}
\end{equation}
where $\vec{\lambda}^\mathrm{prt}$ are the observable polarization parameters
of the promptly produced $S$ state, being $\vec{\lambda} = (\lambda_\vartheta,
\lambda_\varphi, \lambda_{\vartheta \varphi})$, $\vec{\lambda}^\mathrm{dir}$
are the polarization parameters of the directly produced $S$ state, $R(P_1)$
and $R(P_2)$ the fractions of events produced by the decays of $P_1$ and $P_1$
states and $\vec{\lambda}^{P_1}$ and $\vec{\lambda}^{P_2}$ the corresponding
polarizations.

\begin{figure}[tb!]
  \centering
  \includegraphics[width=0.45\textwidth]{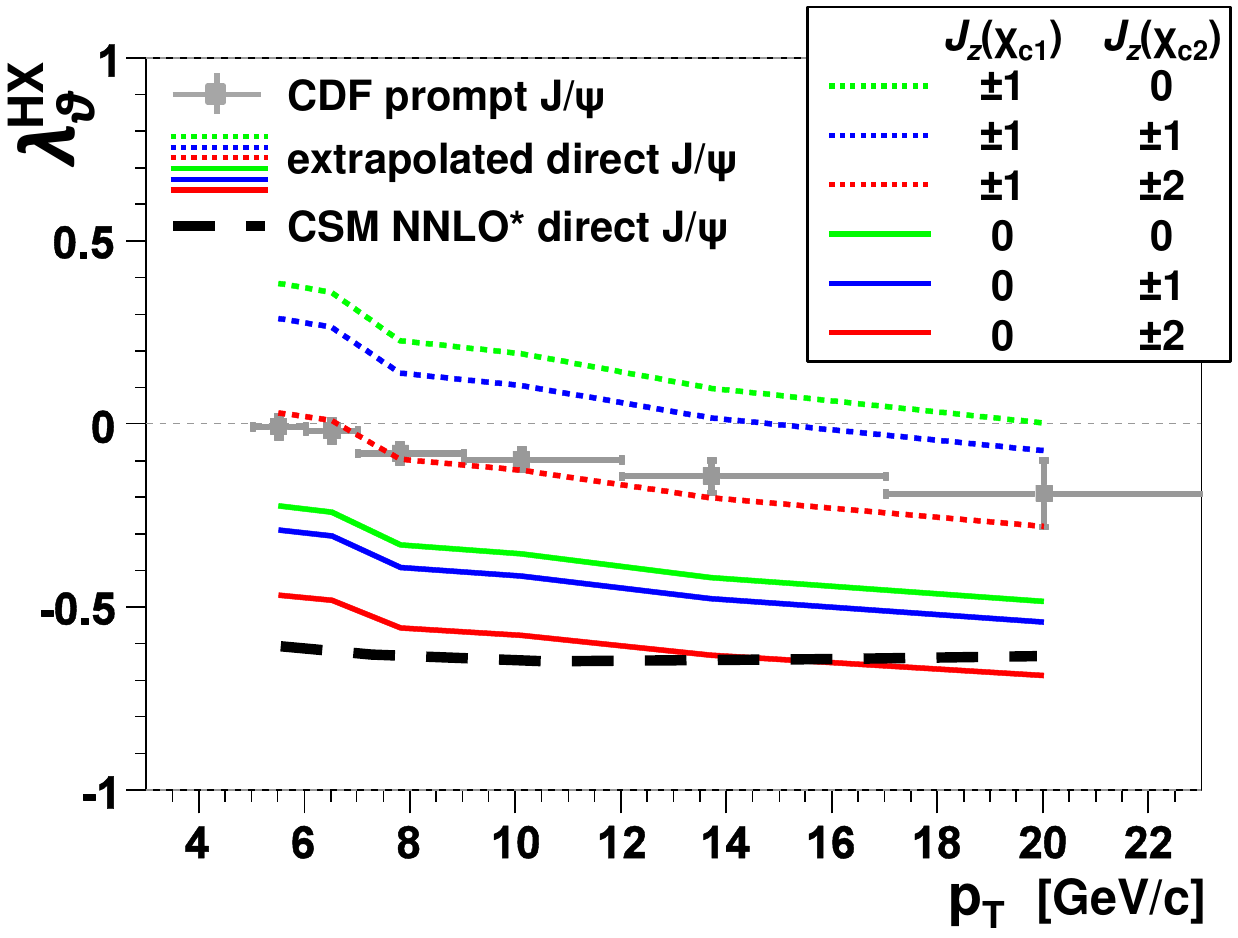}
  \caption{ Direct-J/$\psi$ polarizations ($\lambda_{\vartheta}$) extrapolated
  from the CDF measurement of prompt-J/$\psi$ polarization (in the helicity
  frame), using several scenarios for the $\chi_c$ polarizations. The order of the lines
  from top to bottom is the same as in the legend.
  The NNLO* prediction of direct-J/$\psi$ polarization including only colour-singlet
  contributions~\cite{bib:CSM_directQQpol} is also shown. }
  \label{fig:CDF_direct_extrapolation}
\end{figure}

Figure~\ref{fig:CDF_direct_extrapolation} illustrates how the CDF measurement
of prompt-J/$\psi$ polarization~\cite{bib:CDF_JpsiPol} can be translated in a
range of possible values of the direct-J/$\psi$ polarization, using
Eq.~\ref{eq:addition_rule}, the available information about the feed-down
fractions and all possible combinations of hypotheses of pure polarization
states for $\chi_{c1}$ and $\chi_{c2}$. The feed-down fraction is set to 0.42,
two standard deviations higher than the central CDF value (Eq.~\ref{eq:Rchic});
using 0.30 simply decreases the spread between the curves. The $R(\chi_{c2}) /
R(\chi_{c1})$ ratio is set to 0.40; changes remaining compatible with the CDF
measurement give almost identical curves.

In the scenario in which $\chi_{c1}$ and $\chi_{c2}$ are produced with,
respectively, $J_z = 0$ and $J_z = \pm 2$ polarizations the CDF measurement is
seen to be described by partial next-to-next-to-leading order Colour Singlet
Model predictions (NNLO$^*$ CSM) for directly produced $S$-states
quarkonia~\cite{bib:CSM_directQQpol}. This is the J/$\psi$ polarization
scenario that we will adopt in following considerations. Its validity can be
probed by experiments able to discriminate if the J/$\psi$ is produced together
with a photon such that the two are compatible with being $\chi_{c1}$ or
$\chi_{c2}$ decay products. Such dilepton events, resulting from $\chi_c$
decays, should show a full transverse polarization
($\lambda_\vartheta^{\chi_{c1}} = \lambda_\vartheta^{\chi_{c2}} = +1$), while
the directly produced J/$\psi$ mesons should have a strong longitudinal
polarization ($\lambda_\vartheta^\mathrm{dir} \simeq -0.6$).

We will not discuss here a possible scenario based on non-relativistic QCD
(NRQCD) calculations, which include non-perturbative contributions and, in
particular, colour-octet processes. In fact, the large transverse polarization
predicted by current calculations for the directly produced
$S$-states~\cite{bib:NRQCD_promptQQpol} could be reconciled with the prompt CDF
data only assuming a huge deviation from the measured feed-down from $\chi_c$
[$R(\chi_{c}) \simeq 70\%$] and, at the same time, the maximum possible
longitudinal polarization for the J/$\psi$ from $\chi_c$ (this latter
assumption would be in contradiction with the corresponding prediction of NRQCD
itself).

\begin{figure}[tb!]
  \centering
  \includegraphics[width=0.46\textwidth]{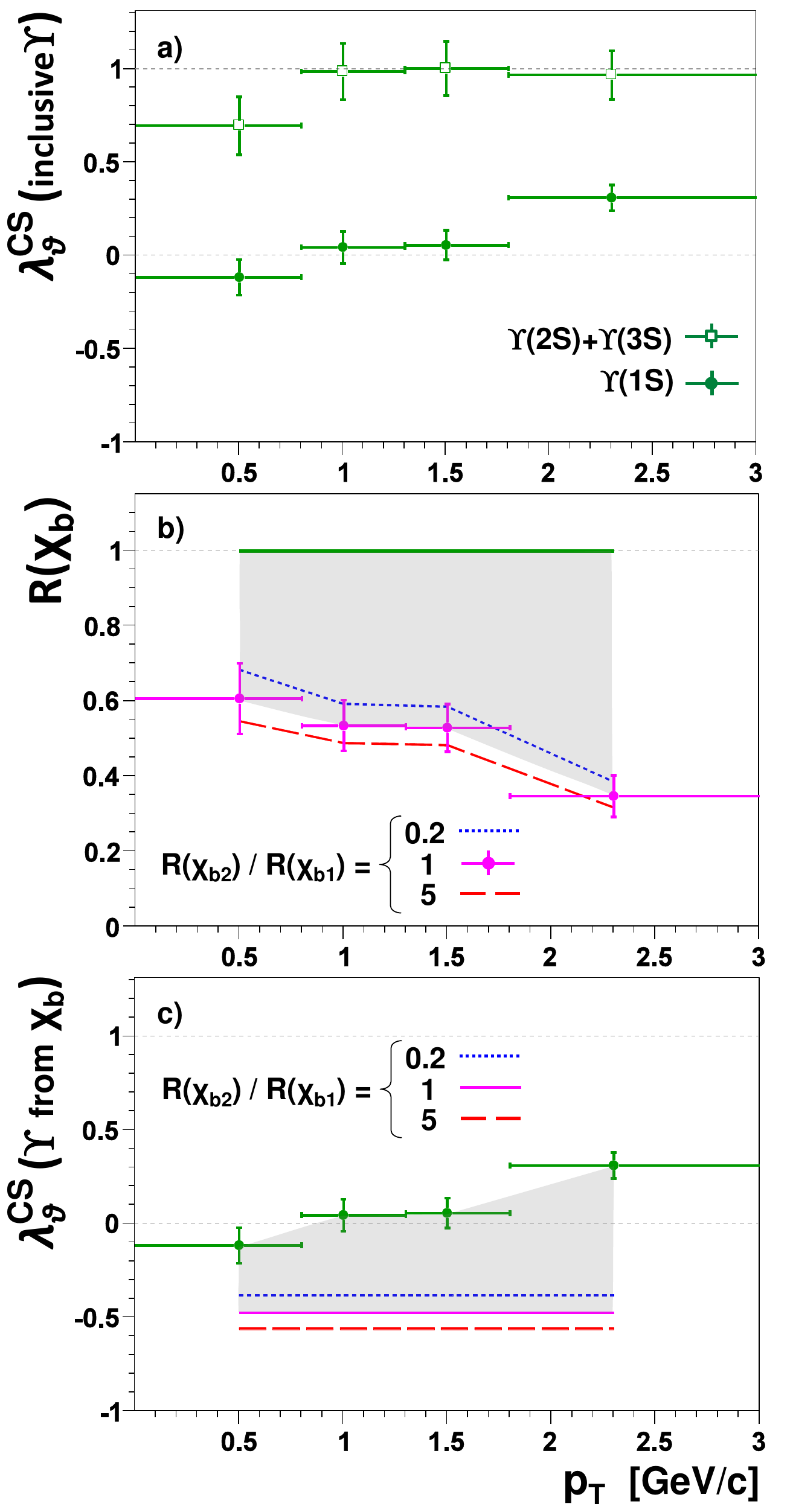}
  \caption{ The E866 measurement of $\Upsilon(1S)$ and $\Upsilon(2S)+\Upsilon(3S)$
  polarizations (in the Collins-Soper frame~\cite{bib:CollinsSoper})
  as a function of $p_{\rm T}$ (a), the deduced ranges for the fraction of $\Upsilon(1S)$
  mesons coming from $\chi_{b}$ decays (b) and the deduced range of their possible polarizations (c).
  An additional systematic uncertainty of $\pm 0.06$ is not included in the error bars
  of the $\lambda_{\vartheta}$ data points in (a).
  The error bars in the derived lower limit of the range for $R(\chi_{b})$ (b) reflect the uncertainty
  in the $\lambda_{\vartheta}$ measurements, assuming that the global systematic uncertainty affects
  the $\Upsilon(1S)$ and $\Upsilon(2S)+\Upsilon(3S)$ measurements in a fully correlated way.
  The lower limits for $R(\chi_{b})$ and $\lambda_{\vartheta}(\Upsilon\mathrm{\:from\:}\chi_{b})$
  depend on the ratio $R(\chi_{b2})/R(\chi_{b1})$, for which three
  different values are assumed. }
  \label{fig:E866_feeddown}
\end{figure}

We will base our second scenario, for the bottomonium family, on the precise
and detailed measurement of E866~\cite{bib:e866_upsilon}, shown in
Fig.~\ref{fig:E866_feeddown}a. This result offers several interesting cues. It
is remarkable that the $\Upsilon(2S)$ and $\Upsilon(3S)$ are found to be almost
fully polarized, while the $\Upsilon(1S)$ is only weakly polarized. The most
reasonable explanation of this fact is that the fraction of $\Upsilon(1S)$
mesons coming from $\chi_{b}$ decays is large and its polarization is very
different with respect to the polarization of the directly produced
$\Upsilon(1S)$. In fact, in the assumption that all directly produced $S$
states have the same polarization, we can translate the E866 measurement into a
lower limit for the feed-down fraction $R(\chi_{b})$ from $P$ states, summing
together $1P_1$, $1P_2$, $2P_1$, $2P_2$ contributions. We use
Eq.~\ref{eq:addition_rule} and the values in Table~\ref{tab:lambda_vs_helicity}
(the average longitudinal momentum of the $\Upsilon(1S)$ in the E866 data is
$\sim 4.5$~GeV$/c$) assuming that the $\Upsilon(2S)+\Upsilon(3S)$ result has a
negligible contamination from $\chi_{b}^\prime \rightarrow \Upsilon(2S) \gamma$
decays and, therefore, provides a good evaluation of the polarization of the
directly produced $S$ states (a conservative assumption for this specific
calculation). The lower limit for $R(\chi_{b})$ corresponds to the case
$J_z(\chi_{b1}) = J_z(\chi_{b1}^\prime) = \pm 1$, $J_z(\chi_{b2}) =
J_z(\chi_{b2}^\prime) = 0$, in which the $\Upsilon(1S)$ mesons from $\chi_{b}$
decays have the largest negative value of $\lambda_{\vartheta}$. The result,
depending slightly on the assumed ratio between $P_2$ and $P_1$ feed-down
contributions, is shown in Fig.~\ref{fig:E866_feeddown}b as a function of
$p_{\rm T}$. More than $50\%$ of the $\Upsilon(1S)$ are produced from $P$
states for $\langle p_{\mathrm{T}} \rangle \simeq 0.5$~GeV$/c$, and more than
$30\%$ for $\langle p_{\mathrm{T}} \rangle \simeq 2.3$~GeV$/c$. These limits
are appreciably higher than the value of the feed-down fraction of J/$\psi$
from $\chi_{c}$ measured at similar energy, low $p_{\rm T}$ and mid
rapidity~\cite{bib:HERAB_chic}. We remind that we have obtained only a lower
limit (no upper limit is implied by the data), corresponding to the case in
which $\chi_{b1}$ and $\chi_{b2}$ are always produced in the same very specific
and pure angular momentum configurations. Any deviation from this extreme case
would lead to higher values of the indirectly determined feed-down fraction.

The E866 measurement data also set an upper limit on the combined polarization
of $\chi_{b1}$ and $\chi_{b2}$. Figure~\ref{fig:E866_feeddown}c shows the
derived range of possible polarizations of $\Upsilon(1S)$ coming from
$\chi_{b}$. The upper bound, corresponding to $R(\chi_{b}) = 1$, coincides with
the measured $\Upsilon(1S)$ polarization. The lower bound, slightly depending
on the relative contribution of $\chi_{b1}$ and $\chi_{b2}$, is not influenced
by the E866 data and corresponds to the minimum ($p_{\mathrm{T}}$ dependent)
value of $R(\chi_{b})$ represented in Fig.~\ref{fig:E866_feeddown}b. The second
strong indication of the E866 data is, therefore, that at low $p_{\rm T}$ the
$\Upsilon(1S)$ coming from $\chi_{b}$ decays has a longitudinal component in
the Collins-Soper frame larger than $\sim 30\%$ ($\lambda_{\vartheta} \lesssim
0.1$), being $\sim 60\%$ ($\lambda_{\vartheta} \sim -0.5$) the maximum amount
of longitudinal polarization that the $\Upsilon(1S)$ produced in this way is
allowed to have.

Also CDF has measured~\cite{bib:upsCDF}, at higher $p_{\rm T}$, an almost
unpolarized production of the $\Upsilon(1S)$ mesons. However, the precision of
the $\Upsilon(2S)$ and $\Upsilon(3S)$ data does not allow us to draw any
conclusion about the difference between the polarizations of directly and
indirectly produced states and, therefore, to infer a possible scenario of
polarizations for the $\Upsilon(1S)$ coming from $\chi_{b}$ decays. A
comparison with the theory predictions for the directly produced
$\Upsilon(1S)$~\cite{bib:CSM_directQQpol,bib:NRQCD_promptQQpol} would lead to
conjectures identical to those made in the J/$\psi$ case, and, therefore, to a
bottomonium polarization scenario completely analogous to the charmonium
scenario described above.

\section{Basic procedures and tools}
\label{sec:formulas}

The E866 example suggests an alternative method to determine the polarization
of the $P$-states, particularly suitable to certain experimental conditions and
always useful as a cross-check of direct determinations. In fact, referring
again to the scenario of Fig.~\ref{fig:E866_feeddown}, a measurement of
$R(\chi_{b})$ would transform the upper bound on the polarization of
$\Upsilon(1S)$ from $\chi_{b}$ decays into a univocal determination. We can
formulate a general way of measuring the combined polarization of $P_1$ and
$P_2$ states, consisting in the following set of measurements: 1) polarization,
$\vec{\lambda}^{1S}$, of the inclusively produced prompt-$1S$ state; 2)
polarization, $\vec{\lambda}^{2S}$, of the $2S$ and/or $3S$ states, assumed to
be mostly directly produced; 3) fraction, $R(P)$, of $1S$ states produced in
the decays of $P$ states. The polarization of the $1S$ states coming from $P$
states can then be determined as using the expression
\begin{equation}
\vec{\lambda}^{P} \, = \, \frac{  (3 + \lambda_\vartheta^{2S})
\vec{\lambda}^{1S} - [1- R(P)] (3 + \lambda_\vartheta^{1S}) \vec{\lambda}^{2S}
} { R(P) ( 3 + \lambda_\vartheta^{1S} ) + \lambda_\vartheta^{2S} -
\lambda_\vartheta^{1S}  }
 \, ,
\label{eq:lambdaP_measurement}
\end{equation}
obviously defined only for $R(P) > 0$ ($\vec{\lambda}^{2S} - \vec{\lambda}^{1S}
\to 0$ for $R(P) \to 0$). As discussed in Ref.~\cite{bib:chi_polarization},
$\vec{\lambda}^{P_1}$ and $\vec{\lambda}^{P_2}$, anisotropy parameters of the
dilepton decay distribution of the daughter $1S$ state, reflect univocally the
average angular momentum configurations in which the $P_1$ and $P_1$ states are
produced. A measurement of $\vec{\lambda}^{P}$, merging $P_1$ and $P_1$
polarizations, can give significant indications, especially if its value is
close to the boundaries of the parameter space and, therefore, does not suffer
cancellation effects (as the E866 example suggests). This method is convenient
if the event sample becomes too small after the requirement of a photon coming
from the $P \to S$ transition, precluding a detailed angular analysis.

For our main result, presented in the next section, we will make use of the
inverse procedure, in which a determination of $R(P)$ is obtained by performing
dilepton polarization measurements. For example, from measurements of
$\lambda_\vartheta^{1S}$, $\lambda_\vartheta^{2S}$ and $\lambda_\vartheta^{P}$
the $\chi$ feed-down is determined as
\begin{equation}
R(P) \, = \, \frac{  (3 + \lambda_\vartheta^{P}) ( \lambda_\vartheta^{2S} -
\lambda_\vartheta^{1S} ) } {(3 + \lambda_\vartheta^{1S}) (
\lambda_\vartheta^{2S} - \lambda_\vartheta^{P} ) } \, .
\label{eq:RP_measurement}
\end{equation}
The significance of this indirect determination will, in general, depend on the
choice of the polarization frame and will be higher in a frame where the
differences between the three $\lambda_\vartheta$ parameters are more
significant. An analysis considering also the azimuthal anisotropy parameters
$\lambda_\varphi$ and $\lambda_{\vartheta\varphi}$ (formulas similar to
Eq.~\ref{eq:RP_measurement}, but depending also on these two parameters, can be
easily deduced from Eq.~\ref{eq:addition_rule}) would lead to the maximum
significance independently of the reference frame. A simpler alternative to
achieve the same result is to use the frame-independent parameter $\mathcal{F}
= (1 + \lambda_\vartheta + 2 \lambda_\varphi)/(3+\lambda_\vartheta)$ introduced
in Ref.~\cite{bib:invariants}:
\begin{equation}
R(P) \, = \, \frac{   \mathcal{F}^{2S} - \mathcal{F}^{1S}  } { \mathcal{F}^{2S}
- \mathcal{F}^{P}  } \, . \label{eq:RP_measurement_F}
\end{equation}

\section{Nuclear dissociation of \boldmath{$P$} states}
\label{sec:suppression}

\begin{figure}[tb!]
  \centering
  \includegraphics[width=0.4\textwidth]{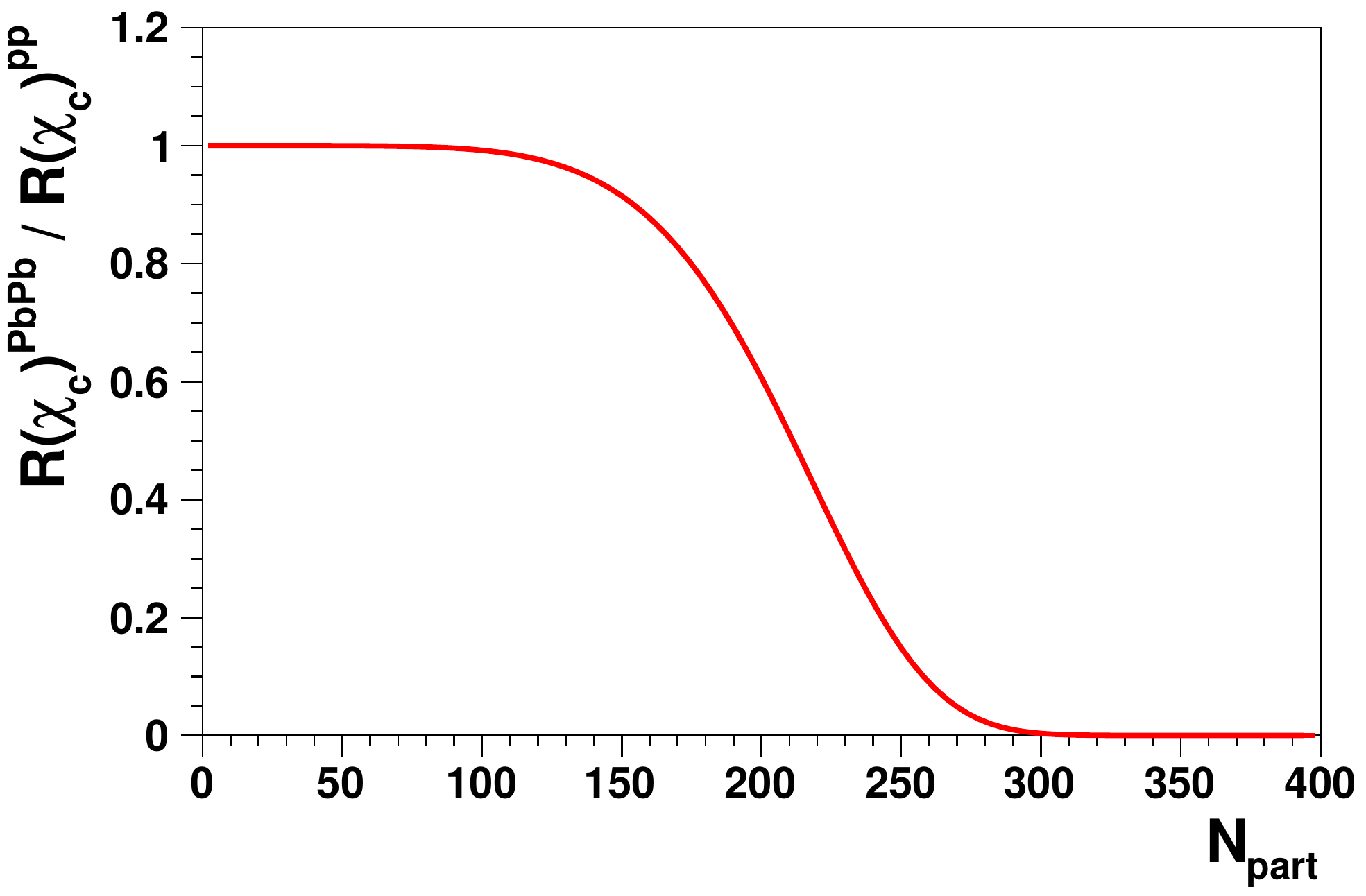}
  \includegraphics[width=0.4\textwidth]{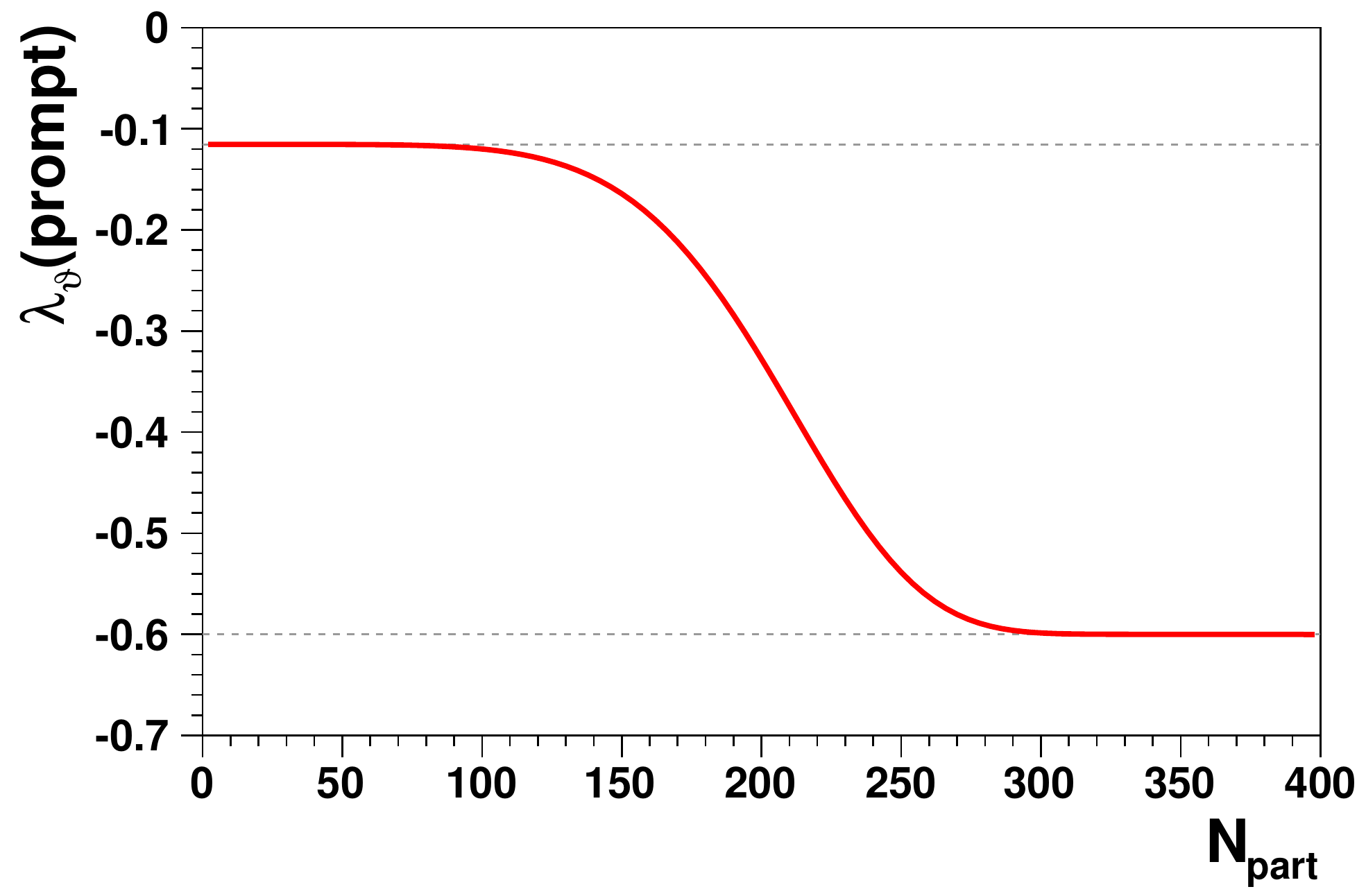}
  \caption{ A hypothetical variation of $R(\chi_c)$ (normalized to the pp value)
  with the centrality of the Pb-Pb collision (top) and the consequent variation
  of the polarization parameter $\lambda_\vartheta$ of the prompt-J/$\psi$ decay distribution
  (bottom), according to the charmonium polarization scenario discussed in the text. }
  \label{fig:pol_seq_suppr}
\end{figure}
\begin{figure}[tb!]
  \centering
  \includegraphics[width=0.4\textwidth]{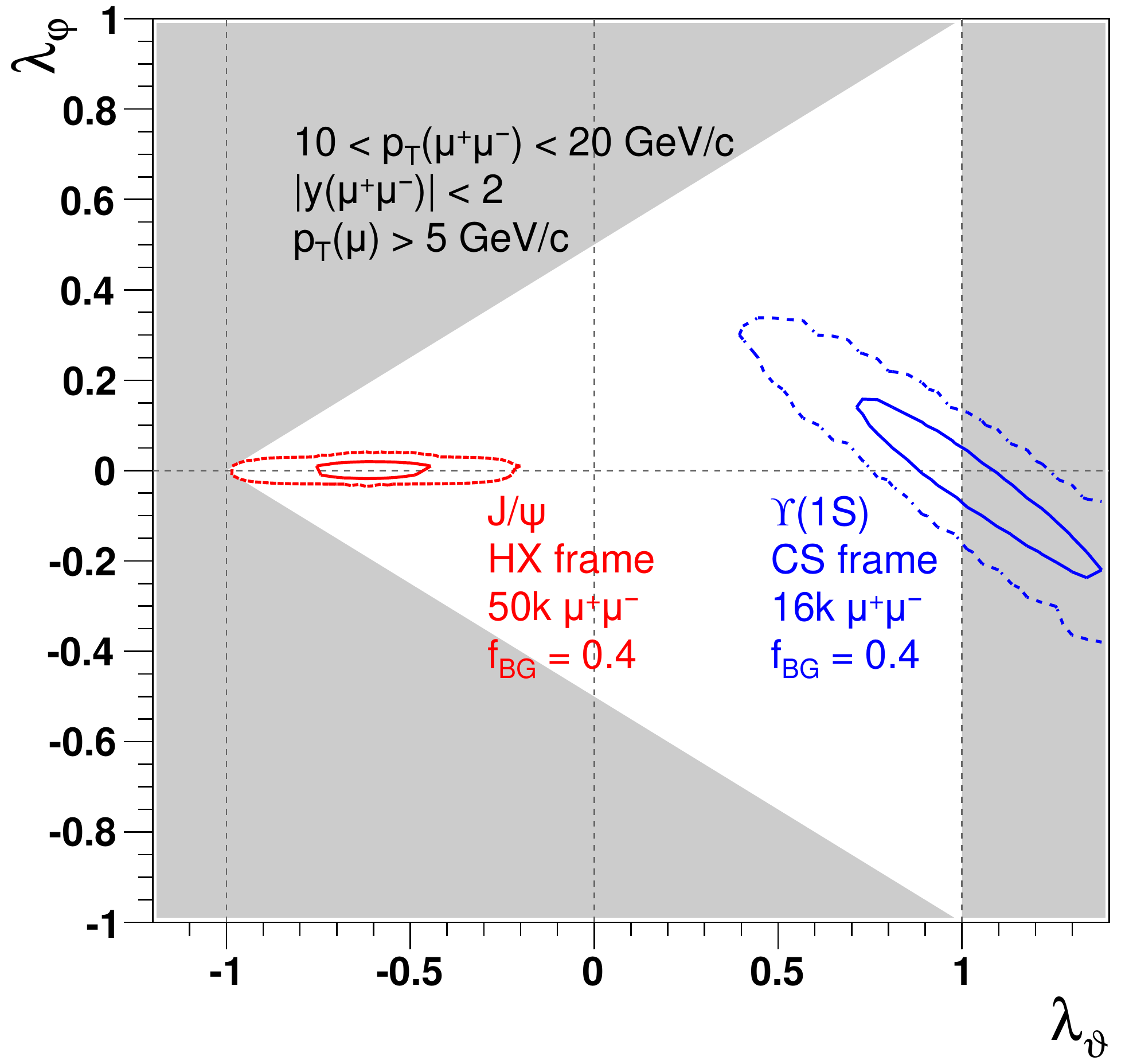}
  \caption{ Results of pseudo-measurements of the prompt-J/$\psi$ and $\Upsilon(1S)$
  dilepton decay distributions in Pb-Pb collisions, according to the corresponding scenarios
  discussed in the text. The background fraction ($f_{BG}$) is defined within $\pm 2 \sigma$
  from the maximum of the invariant mass distribution, where $\sigma$, the experimental
  width, is, respectively, $4$ and $10$~MeV$/c^2$.
  The contours represent $68$ and $99\%$ confidence levels.
  The shaded area is physically forbidden. }
  \label{fig:pol_measurements}
\end{figure}

The possibility of determining the feed-down fraction from $P$ states purely on
the basis of \emph{dilepton} properties is particularly valuable in the
perspective of quarkonium measurements in heavy-ion collisions, where a direct
determination of the $\chi$ yields is essentially impossible.

Figure~\ref{fig:pol_seq_suppr} illustrates the concept of the method. The top
panel shows an hypothetical $R(\chi_c)$ pattern inspired from the sequential
charmonium suppression scenario, in which the $\chi_c$ yield disappears rapidly
beyond a critical value of the number of nucleons participating in the
interaction ($N_{\rm part}$). This effect would be reflected by a change in the
observed prompt-J/$\psi$ polarization. As shown in the bottom panel, according
to the scenario presented in Sec.~\ref{sec:scenarios} the polarization should
become significantly more longitudinal (in the helicity frame) after the
disappearance of the transversely polarized feed-down contribution due to
$\chi_c$ decays. We are assuming that the ``base'' polarizations of the
directly produced $S$ and $P$ states remain essentially unaffected by the
nuclear medium and are, therefore, not distinguishable from those measurable in
pp collisions.

In general, the determination of $R(\chi_c)$ as a function of number of
participants from the corresponding polarization measurements requires the
knowledge of the polarization of the J/$\psi$ coming from $\chi_c$ decays
(Eq.~\ref{eq:RP_measurement}). This measurement can be made in pp collisions,
merging the $\chi_{c1}$ and $\chi_{c2}$ contributions and using the simplified
procedure illustrated in Sec.~\ref{sec:formulas}
(Eq.~\ref{eq:lambdaP_measurement}).

As a simpler option, a test of the sequential suppression pattern can be made
by comparing the prompt-J/$\psi$ polarization measured in pp collisions (or
peripheral nucleus-nucleus) with the one measured in central nucleus-nucleus
collisions and checking that this latter tends to the polarization of the
$\psi^\prime$, also measured in pp collisions.

In this illustration we have neglected the role of the $\psi^\prime$
suppression, which, in the sequential suppression scenario, would lead to a
slight increase of $R(\chi_c)$ (and, therefore, to a slight reduction of the
prompt-J/$\psi$ polarization) \emph{before} the $\chi_c$ disappearance. This
would make the change in polarization due to $\chi_c$ suppression more drastic
as a function of $N_{\rm part}$, but would not modify the difference between
the prompt-J/$\psi$ polarizations observed in proton-proton and central
nucleus-nucleus collisions.

Moreover, we are assuming that the parton recombination into low-$p_{\rm T}$
J/$\psi$ in central collisions does not play a role. For this reason, in what
follows we will consider relatively high-$p_{\rm T}$ measurements. However, we
remark that recombination would probably change the above picture in a
distinctive way, leading to a second observable change in the J/$\psi$
polarization as a function of centrality.

The same method can be applied to the measurement of $\chi_b$ suppression using
$\Upsilon(1S)$ polarization. According to the scenario based on the E866 data
(Sec.~\ref{sec:scenarios}), in pp (and peripheral Pb-Pb) collisions the
$\Upsilon(1S)$ should be only slightly polarized, reflecting the mixture of
directly and indirectly produced states with opposite polarizations. In central
Pb-Pb collisions the $\Upsilon(1S)$ would acquire the fully transverse
polarization characteristic of the directly produced $S$ states, indicating the
suppression of the $P$ states. Also in this case the simple stepwise behaviour
as a function of $N_{\rm part}$ would be slightly contaminated, but not made
less visible, by the suppression of $\Upsilon(2S)$ and $\Upsilon(3S)$. On the
other hand, the presence of the $2P$ states in the bottomonium family would add
an intermediate step in the pattern of polarization change.

Figure~\ref{fig:pol_measurements} shows the results of pseudo-measurements of
the prompt-J/$\psi$ and $\Upsilon(1S)$ polarizations in central Pb-Pb
collisions, based on, respectively, about $30$k and $10$k reconstructed signal
events in the dimuon channel and assuming in both cases a background fraction
of $40\%$. The dimuon $p_{\rm T}$ and rapidity are in the ranges $10 < p_{\rm
T} < 20$~GeV$/c$ and $|y| < 2$. Only events where both muons have $p_{\rm T} >
5$~GeV$/c$ are included in the reconstruction, in order to simulate the typical
reduction of angular phase-space affecting this kind of measurements at the
LHC. The central values of the measurements correspond to the expected
polarizations, strongly longitudinal and fully transverse in the two respective
scenarios, after the melting of the $\chi$ states. The results exclude large
part of the $(\lambda_\vartheta, \lambda_\varphi)$ plane and, in particular,
the region around the origin, containing the (precisely determined) pp values
($\lambda_\vartheta \sim -0.15$ in the J/$\psi$ scenario and $\lambda_\vartheta
\sim 0.3$ in the $\Upsilon(1S)$ scenario, assuming $R(\chi_{b}) \simeq 0.4$).
Such measurements would represent a significant indication of the nuclear
disassociation of the $\chi$ states.

\section{Summary}
\label{sec:summary}

We have demonstrated that it is possible to determine the nuclear suppression
of the $\chi_c$ and $\chi_b$ states through measurements of the inclusive
dilepton decay distributions of prompt J/$\psi$ and $\Upsilon(1S)$.

In a preliminary discussion we have illustrated how the polarizations of the
directly produced $S$- and $P$-state quarkonia are likely to differ
significantly from one another. Given that the feed-down contributions of $P$
states to the prompt $S$ states are large, this means that there must be a
measurable difference between the decay distributions of indirectly and
directly produced $S$ states. The hypothesis is strongly supported by the E866
$\Upsilon$ data and is seen to reconcile the CDF prompt-J/$\psi$ and
$\Upsilon(1S)$ measurements with perturbative-QCD predictions for the
polarizations of the directly produced states. These are interesting
indications for the understanding of quarkonium production and should be
verified with detailed polarization measurements distinguishing between the
properties of directly and indirectly produced states. We have also proposed an
alternative and simplified way of determining the polarizations of J/$\psi$ and
$\Upsilon(1S)$ coming from decays of $P$ states, using measurements of $R(P)$
and of the polarizations of the $1S$ and $2/3S$ states, instead of studying
directly the angular distribution of events identified with the presence of a
radiated photon.

With the above premises, we have shown that a change in the relative yield of
$S$ and $P$ states from proton-proton to nucleus-nucleus collisions is directly
reflected in an observable change of the prompt-J/$\psi$ [or $\Upsilon(1S)$]
polarization. The sequential dissociation scenario has a particularly clean
polarization signature: the melting of the $\chi$ states would be signalled by
the observation of a significantly larger prompt-J/$\psi$ [$\Upsilon(1S)$]
polarization than the one measured in pp collision. After the complete
suppression of $\chi$ production, the polarization should approach the one
measured (in pp collisions) for the $\psi^\prime$ [$\Upsilon(2/3S)$].

In conclusion, quarkonium polarization can be used as a new probe for the
formation of a deconfined medium. This method, based on the study of dilepton
kinematics alone, provides a feasible and clean alternative to the direct
measurement of the $\chi$ yields through reconstruction of radiative decays.
With sizeable J/$\psi$ and $\Upsilon(1S)$ event samples to be collected in
nucleus-nucleus collisions, the LHC experiments have the potential to provide a
clear insight into the role of the $\chi$ states in the dissociation of
quarkonia, taking a crucial step forward in establishing the validity of the
sequential melting mechanism.

\bigskip

We acknowledge support from Funda\c{c}\~ao para a Ci\^encia e a Tecnologia,
Portugal, under contracts SFRH/BPD/42343/2007, CERN/FP/116367/2010 and
CERN/FP/116379/2010. We also acknowledge interesting discussions with C.
Louren\c{c}o.


\end{document}